\documentclass{elsart}
\usepackage{latexsym}
 
\begin{document}
\begin{frontmatter}
\journal{J. Mol. Struct. (Theochem)}
\date{Mar 18, 1997}
 \title{The size-extensitivity of correlation energy estimators based on
 effective characteristic polynomials\thanksref{ECCC3}}
 \author{Herbert H. H. Homeier\thanksref{HHHH}}
\address{
 Institut f\"ur Physikalische und Theoretische Chemie,
  Universit\"at Regensburg,
   D-93040 Regensburg, Germany}
   \thanks[HHHH]{E-mail: na.hhomeier@na-net.ornl.gov,\\ 
   WWW: http://www.chemie.uni-regensburg.de/\%7Ehoh05008/}
   \thanks[ECCC3]{Paper 27 at the 3$^{rd}$ Electronic Computational
   Chemistry Conference, 1996. \\ Regensburg Preprint TC-QM-96-3.
   {}\quad\hfill Typeset using elsart.cls}
\begin{abstract}
Estimators $\Pi n$ for the correlation energy can be computed as roots  of
effective characteristic polynomials of degree $n$. The coefficients of these
polynomials are derived from the terms of the perturbation series of
the energy. From a fourth-order M{\o}ller-Plesset (MP4) calculation one
can calculate with negligible effort a size-extensive estimator $\Pi 2$
that is in many cases much closer to the full CI correlation energy of the ground state than the
MP4 value. [H.H.H. Homeier, J.\ Mol.\ Struct.\ (Theochem)  366, 161 (1996)%
]
Here, we prove that the estimators $\Pi n$ for $n>2$ are size-extensive
if they are calculated from the MP series.

\end{abstract}
\end{frontmatter}

 \section{Introduction}
 \label{secint}

Many-body perturbation theory is a convenient tool to estimate the
correlation energy of molecular systems. Usually, one calculates an
estimate for the correlation energy by term-by-term summation of the
M{\o}ller-Plesset (MP) series
\begin{equation}\label{eq0}
E = E_{0} + E_{1} + E_{2} + E_{3} +  E_{4} +  E_{5} + \dots\>,
\end{equation}
as a partial sum
\begin{equation}\label{eq0a}
E^{(n)} = \sum_{j=0}^{n} E_j\>
\end{equation}
that is usually denoted by MP$n$.
But this approximate value does not optimally exploit the information content of the
terms $E_j$ of the MP series. Better estimates can be obtained by
using convergence acceleration or extrapolation methods to sum the
perturbation series, as for instance Pad{\'e} approximants or methods
based on effective characteristic polynomials. The method of effective
characteristic polynomials has been introduced by {{\v C}{\'\i}{\v
z}ek} and coworkers recently. It has been applied to the summation of the
divergent perturbation series of anharmonic oscillators
\cite{CizekWenigerBrackenSpirko96} and to correlation energies of model
systems \cite{Bracken94,BrackenCizek94,%
BrackenCizek95a,BrackenCizek95b,CizekBracken95,%
DowningMichlCizekPaldus79,TakahashiBrackenCizekPaldus95}. Extensions of
the method for the simultaneous treatment of several perturbation
series have recently been proposed \cite{Homeier96Hab} but this will not be
considered in the following.

If only perturbation energies up to the fourth order are available then
one can use the $\Pi 2$ estimator \cite{Homeier96}. The $\Pi 2$ estimator
is obtained as a root of a second-degree effective characteristic
polynomial. The coefficients of this polynomial are related to the
terms of the perturbation series. The $\Pi 2$ estimator can be
calculated easily from the terms $E_0,\dots,E_4$ of the MP series and
is size-extensive. As is well-known, the latter property is important
for the treatment of larger systems. For a series of benchmark systems
the $\Pi 2$ estimator proved to be relatively accurate as compared to a
number of other estimators \cite{Homeier96}.

Analogous estimators $\Pi n$ can be derived from effective
characteristic polynomials of higher degrees $n$. In the present
contribution, we sketch the method of effective characteristic
polynomials, basically to fix notation. Then, it is proved that the
estimators $\Pi n$ for $n>2$ are also size-extensive if the underlying
perturbation theory is size-extensive, as is the case for the MP series.

 \section{Correlation energy estimators based on effective
 characteristic polynomials}
 \label{seccor}
  In this section, we sketch the method of the effective
  characteristic polynomials. The characteristic polynomial
  $P_n(E)$ of degree $n$ in the unknown energy $E$  has in the
  linear variation method the form
\begin{equation}\label{eqC1}
P_n(E)={\rm det}\,\left\vert \left\langle \phi_j \vert H \vert
\phi_k \right\rangle -E\,\delta_{j,k}\right\vert \>
\end{equation}
  where $\{\phi_j\}_{j=1}^{n}$ are $n$ orthonormal basis
  functions, and $H$ is the  Hamiltonian.   If the Hamiltonian
  can be written as $H=H_0+\beta V$, the polynomial has the form
  (\cite{CizekWenigerBrackenSpirko96}, Eq. (3.2))
\begin{equation}\label{eqC2}
P_n(E) = \sum_{j=0}^{n} E^j \,\sum_{k=0}^{n-j} f_{n,j,k} \beta^k
\end{equation}
  with $f_{n,n,0}=1$. This leaves $L+1$ coefficients
  $f_{n,j,k}$ to be determined where $L=n(n+3)/2-1$. In the method of the
  characteristic polynomial, they are obtained from the
  coefficients $E_j$ of the perturbation series for $E$
\begin{equation}\label{eqC3}
E = \sum_{j=0}^{\infty} E_j \, \beta^j\>.
\end{equation}
  Since $P_n(E)=0$ for an eigenvalue $E$, one demands
\begin{equation}
P_n(E_0+\beta E_1 + \beta^2 E_2 + \dots ) = O(\beta^{L+1})\>.
\end{equation}
  This means that the first ${L+1}$ coeffients of the Taylor
  series (in $\beta$) of the left-hand side of this equation,
  i.e. up to the coefficient of $\beta^{L}$,  all have to
  vanish. This yields a linear equation system for the unknown
  $f_{n,j,k}$. We assume in the following that this linear
  equation system yields a unique solution for the coefficients
  $f_{n,j,k}$. After the determination of the coefficients, the
  effective characteristic polynomial is fixed and also denoted
  by $P_n(E)=P_n[E_0,\dots,E_{L}](E)$ in order to make the
  dependence on the terms $E_j$ explicit. The effective
  characteristic equation $P_n[E_0,\dots,E_{L}](E)=0$ may then
  be solved for $E$. The lowest root is called $\Pi
  n[E_0,\dots,E_{L}]$ or, more simply, $\Pi n$, if the values of
  the $E_j$ are plain from the context.

  In the case $n=2$ one obtains for $\beta=1$ the simple expression
\begin{eqnarray}\label{eqP2}
 \Pi 2 &=& E_0 + E_1
       +
 {\displaystyle \frac {E_2^2}{2}}
 \,\frac{\displaystyle
    E_2 -E_3
   }
   {\displaystyle  E_2\,E_4 - E_3^2
   }  \nonumber \\
    &+& {\displaystyle \frac {E_2^2}{2}}
 \,\left[\frac{\displaystyle
(E_2- E_3)^2 - 4 \,(E_2\,E_4- E_3^2)
   }
   {\displaystyle  (E_2\,E_4 - E_3^2)^2
   }\right]^{{1/2}}\>.
\end{eqnarray}

 \section{Proof of the size-extensitivity}
 \label{secpro}
 Consider a supersystem composed of $M$ identical,
 non-interacting subsystems. Then, the true energy of the
 supersystem is the $M$-fold of the energy of a single
 subsystem. An approximate method for the computation of the
 energy is called size-extensive if the approximate energy for
 the supersystem is the $M$-fold of the approximate energy of a
 single subsystem.

 It is well-known that MP perturbation theory is size-extensive
 order by order. This means that for all $k\ge 0$ the $k$-th
 order term of the perturbation series for the supersystem
 equals $M \, E_k$ if the $k$-th order term of the perturbation
 series of a single subsystem is $E_k$. This implies that
 the MP$k$ estimator is size-extensive for each order $k$.

 In order to prove size-extensitivity of the $\Pi n$ estimators
 with $n>2$, one thus has to show that for each $n$ and $M$ the
 equation
\begin{equation}\label{toprove}
\Pi n[M\,E_0,\dots,M\,E_{L}] = M \,\Pi n[E_0,\dots,E_{L}]
\end{equation}
 holds where, as before, ${L}=n(n+3)/2-1$.

 For given $M$ and $n$ and for given $E_0,\dots,E_{L}$ and
 $\beta$, we can consider the $f_{n,j,k}$ and hence, also the
 effective characteristic polynomial
\begin{equation}
P_n[E_0,\dots,E_{L}](E)=\sum_{j=0}^{n} E^j \,\sum_{k=0}^{n-j} f_{n,j,k} \beta^k
\end{equation}
 as known.
 We introduce a new polynomial by
\begin{eqnarray}
\tilde P_n(E) &=& M^n \, P_n[E_0,\dots,E_{L}](E/M) \label{newpola}\\
              &=& \sum_{j=0}^{n} E^j \,\sum_{k=0}^{n-j}
              [f_{n,j,k} M^{n-j}] \beta^k \label{newpolb}\>.
\end{eqnarray}
 Eq.\ (\ref{newpola}) defines the polynomial, while Eq.\
 (\ref{newpolb}) shows that $\tilde P_n(E)$ can also be regarded
 as an effective characteristic polynomial with new coefficients
 $\tilde f_{n,j,k}= f_{n,j,k} M^{n-j}$. Note that $\tilde
 f_{n,n,0}=1$ holds as required.

 Since
\begin{eqnarray}
&\tilde P_n&(M\, E_0+\beta \,M\, E_1 + \beta^2 M \,E_2 + \dots )
\nonumber \\
&=& M^n\, P_n[E_0,\dots,E_{L}](E_0+\beta \,E_1 + \beta^2 E_2 + \dots )
\nonumber \\
 &=& O(\beta^{L+1})\>,
\end{eqnarray}
 the polynomial $\tilde P_n(E)$ is identical to the effective characteristic
 polynomial for the energies $M\,E_0,\dots,M\,E_L$, i.e.
\begin{equation}
M^n \, P_n[E_0,\dots,E_{L}](E/M) = P_n[M\,E_0,\dots,M\,E_{L}](E)\>.
\end{equation}
 Thus, the complete pattern of roots  is scaled by $M$, since if
 $\epsilon$ is any root of $P_n[E_0,\dots,E_{L}](E)$ then
 $M\,\epsilon$ is a root of $P_n[M\,E_0,\dots,M\,E_{L}](E)$. But
 this proves Eq.\ (\ref{toprove}) since $\Pi n[E_0,\dots,E_{L}]$
 is a root of $P_n[E_0,\dots,E_{L}](E)$.

\section{Concluding Remarks}
Size-extensitivity is an important property that is helpful for obtaining 
reliable energy estimates for larger systems. It is hoped that correlation energy
estimates on the basis of effective characteristic polynomials become more
widespread since they offer a conceptually and computationally relatively simple 
but accurate computational tool as demonstrated in \cite{Homeier96}. 
Only the MP series has to be computed, and its terms
be combined to obtain an effective characteristic polynomial and the correlation energy 
as one of its roots as in Eq.\ (\ref{eqP2}). Some error control is possible by
comparison to some other methods for convergence acceleration of the perturbation series
\cite{Homeier96}. 

An extension of the method  was proposed by the author in 
\cite{Homeier96Hab}. It allows to combine  
information from  the terms of several short perturbation expansions for a small number of 
states.
These simultaneous perturbation series can also be used  
for the construction of a characteristic polynomial. In this way, the somewhat demanding
relation between the degree of the characteristic polynomial and the maximal order of the
perturbation calculation is largely avoided. Applications of this extended method
and a proof of its size-extensitivity are currently under investigation.

\begin{ack}
 The author is pleased to acknowledge  helpful discussions regarding
 the effective characteristic polynomial method with Prof.\ Dr.\ J.\ {\v C}{\'\i}{\v z}ek, 
Prof.\ Dr.\ E.\ J.\ Weniger, and Dr.\ H.\ Mei{\ss}ner. 
The author is grateful to Prof.\ Dr.\ E.\ O.\ Steinborn for his support and the excellent 
working conditions at Regensburg.
\end{ack}

 \end{document}